\documentclass[pra,aps,groupaddress,twocolumns,showpacs]{revtex4}
\usepackage{epsfig,amsmath,graphicx,amssymb,overpic}
\usepackage{bm}
\usepackage{mathrsfs}
\usepackage{graphicx}
\usepackage{epsfig}
\usepackage{bm}
\usepackage{amsmath,amssymb,amsthm}
\usepackage[colorlinks=true,linkcolor=blue]{hyperref}
\usepackage{subfigure}
\usepackage{booktabs}
\usepackage[mathscr]{euscript}
\usepackage{ulem}

\newcommand{\bea}{\begin{eqnarray}}
\newcommand{\eea}{\end{eqnarray}}
\newcommand{\beq}{\begin{equation}}
\newcommand{\eeq}{\end{equation}}

\def\/{\over}

\begin{document}

\title{Searching for missing D'Alembert waves in nonlinear system: Nizhnik-Novikov-Veselov equation}
\author{Man JIA and S. Y. LOU\thanks{Email:lousenyue@nbu.edu.cn}\\
\footnotesize \it Laboratory of Clean Energy Storage and Conversion, \\
\footnotesize \it School of Physical Science and Technology, Ningbo University, Ningbo 315211, P. R. China}

\begin{abstract}
In linear science, the wave motion equation with general D'Alembert wave solutions is one of the fundamental models. The D'Alembert wave is an arbitrary travelling wave moving along one direction under a fixed model (material) dependent velocity. However, the D'Alembert waves are missed when nonlinear effects are introduced to wave motions. In this paper, we study the possible travelling wave solutions, multiple soliton solutions and soliton molecules for a special (2+1)-dimensional Koteweg-de Vries (KdV) equation, the so-called Nizhnik-Novikov-Veselov (NNV) equation. The missed D'Alembert wave is re-discovered from the NNV equation. By using the velocity resonance mechanism, the soliton molecules are found to be closely related to D'Alembert waves. In fact, the soliton molecules of the NNV equation can be viewed as special D'Alembert waves. The interaction solutions among special D'Alembert type waves ($n$-soliton molecules and soliton-solitoff molecules) and solitons are also discussed.
\end{abstract}

\pacs{05.45.Yv,02.30.Ik,47.20.Ky,52.35.Mw,52.35.Sb}
\maketitle

The wave motion equation
\begin{equation}
u_{tt}-c^2u_{xx}=0\label{WME}
\end{equation}
is one of the most fundamental equations in linear physics. The general solutions of \eqref{WME} possess the D'Alembert wave form
\begin{equation}
u=f(\xi)+g(\eta),\ \xi\equiv x-ct,\ \eta\equiv x+ct, \label{DLBE}
\end{equation}
where $f$ and $g$ are arbitrary second-order differentiable functions of the traveling wave variables $\xi$ and $\eta$, respectively. The model parameter $c$, the velocity of the D'Alembert waves, is fixed and related to concrete physical problems. For instance, $c$ is the light velocity in optics and electromagnetics and sound velocity in acoustics and elasticity mechanics.

The simplest nonlinear extension of the wave motion equation is the
Korteweg-de Vries (KdV) equation \cite{KdV}
\begin{equation}
u_t + u_{xxx} + 6uu_x = 0, \label{KdV}
\end{equation}
which
possesses various applications in physics and other scientific fields. It approximately describes the evolution of long, one-dimensional waves in many physical settings, including shallow-water waves with weakly nonlinear restoring forces, long internal waves in a density-stratified ocean, ion acoustic waves in a plasma, acoustic waves on a crystal lattice \cite{KdV} and the 2-dimensional quantum gravity \cite{Guo}. The KdV equation can be solved by using the inverse
scattering transform \cite{IST} and other powerful methods such as the Hirota's bilinear method \cite{Hirota} and the Darboux transformation \cite{Gu}.

The KdV equation and other nonlinear extensions lost the D'Alembert type wave solutions. A natural and important question is can we find possible nonlinear wave extensions such that the missing D'Alembert type waves are still allowed?

In this letter, we study some special types of exact solutions for a special (2+1)-dimensional extension of the KdV equation, the  so-called Nizhnik-Novikov-Veselov (NNV)
equation\cite{NNV1,NNV2,NNV3}
\begin{equation}
u_t+cu_x+du_y+au_{xxx}+bu_{yyy}+3a\left(uv\right)_x+3b\left(uw\right)_y=0,\ u_x=v_y,\ u_y=w_x. \label{NNV}
\end{equation}
It is clear that when $y=x$ and $v=w=u$, the NNV equation \eqref{NNV} is reduced to the KdV equation \eqref{KdV} after some scaling and Galileo transformations.

 The study of soliton molecules (SMs), soliton bound states, is one of the hot topics because some types of SMs have been observed and applied in several physical areas such as the optics \cite{SP05,HK17,LXM,Nano,SA} and Bose-Einstein condensates \cite{LNS}. Some theoretical proposals to form soliton molecules have been established \cite{CK03,YB11}. In Ref. \cite{YanZY}, the similar soliton molecules were numerically verified to exist in the nonlinear dispersive NLS(n,n) equation. Especially, in Refs. \cite{LouS1,LouS2}, a new mechanism, the velocity resonance, to form soliton molecules is proposed. It is found that the standard (1+1)-dimensional KdV equation \eqref{KdV} does not possess soliton molecules. However, in real physics, higher order effects such as the higher order dispersions and higher order nonlinearities have been neglected when the KdV equation \eqref{KdV} is derived \cite{KdV5}. Whence the higher order effects are considered to the usual KdV equation, one can really find some types of SMs \cite{LouS1}. By using the velocity resonance mechanism, some authors find new types of SMs such as the dromion molecules and half periodic kink (HPK) molecules for some other physical systems \cite{LiB,XYT,YZW,KP34}.

To search for the possible D'Alembert type wave solutions, we study the traveling wave solutions of the NNV equation \eqref{NNV} in the form
\begin{equation}
u=U(\tau),\ v=V(\tau),\ w=W(\tau),\ \tau=kx+py+\omega t.\label{UVW}
\end{equation}
Substituting \eqref{UVW} into \eqref{NNV} yields
\begin{equation}
(\omega +ck+dp)U_\tau+(ak^3+bp^3)U_{\tau\tau\tau} +3ak\left(UV\right)_{\tau}+3bp\left(UW\right)_{\tau}=0,\
kU_{\tau}=pV_{\tau},\ pU_{\tau}=kW_{\tau}, \label{NNVtau}
\end{equation}
say,
\begin{eqnarray}
&&V=\frac{k}{p}U, \ W=\frac{ p}kU, \label{NNVVW}\\
&&(\omega +ck+dp)U_\tau+(ak^3+bp^3)U_{\tau\tau\tau}+3\frac{ak^3+bp^3}{pk}\left(U^2\right)_{\tau}=0. \label{NNVU}
\end{eqnarray}
From \eqref{NNVU} we know that when setting
\begin{eqnarray}
p = -\sqrt[3]{\frac{a}b}k,\ \omega=-ck-dp,
\label{NNVpo}
\end{eqnarray}
the travelling wave $U$ becomes an arbitrary D'Alembert type wave in the form
\begin{eqnarray}
u=U\left(\tau\right),\ \tau=x-\sqrt[3]{\frac{a}b}y-\left(c-d\sqrt[3]{\frac{a}b}\right)t, \label{tau1}
\end{eqnarray}
and moves along the direction
$\sqrt[3]{\frac{a}b}x+y$ (perpendicular to $x-\sqrt[3]{\frac{a}b}y$) with the model parameter ($\{a,\ b,\ c,\ d\}$) dependent velocity 
$$\vec{c}=\{c_x,\ c_y\}\equiv \left\{c-d\sqrt[3]{\frac{a}b},\ -\left(c-d\sqrt[3]{\frac{a}b}\right)\sqrt[3]{\frac{b}a}\right\},\  
|c|=\sqrt{c_x^2+c_y^2}=\left|d-c\sqrt[3]{\frac{b}a}\right|\sqrt{1+\left(\frac{a}b\right)^{2/3}}.$$
Because of the arbitrariness of the D'Alembert type wave \eqref{tau1}, the solitary waves and soliton molecules possess quite free shapes.

Fig. 1 shows some special structures which may be used to describe real nonlinear phenomena. Fig. 1a is a kink molecule which is obtained by taking
\begin{eqnarray}
U(\tau)=2 \left[\ln(1+\mbox{\rm e}^{k_1(\tau+\tau_{10})}
+\mbox{\rm e}^{k_2(\tau+\tau_{20})}+a_{12}\mbox{\rm e}^{k_1(\tau+\tau_{10})+k_2(\tau+\tau_{20})}
)\right]_{\tau}
\label{F1a}
\end{eqnarray}
with $a=b=c=d=k_1=a_{12}=1,$ $\tau_{20}=-\tau_{10}=5$ and $k_2=0.8$ at time $t=0$.

Fig. 1b displays the structure of HPK-kink molecule described by
\begin{eqnarray}
U(\tau)=2 \left\{\ln\left[1+(1-0.1\cos(0.5\tau))\mbox{\rm e}^{k_1(\tau+\tau_{10})}
+\mbox{\rm e}^{k_2(\tau+\tau_{20})}
+(1-0.1\cos(1.5\tau))\mbox{\rm e}^{2k_1(\tau+\tau_{10})}\right]\right\}_{\tau}
\label{F1b}
\end{eqnarray}
with $a=b=c=d=k_1=1,$ $\tau_{20}=-\tau_{10}=6$ and $k_2=-0.8$ at time $t=0$.

Fig.1c is a plot of a periodic kink structure expressed by
\begin{eqnarray}
U(\tau)=\tanh\left\{\tau[1-0.5\cos(0.1\tau)]\right\}[1-0.1\cos(2\tau)]
\label{F1c}
\end{eqnarray}
with $a=b=c=d=1$  at time $t=0$.

Fig.1d displays a three-soliton molecule presented by
\begin{eqnarray}
U(\tau)=\frac{6}{(c_1-\tau^2)^2}\exp\left[-\frac{2}{(c_1-\tau^2)^2}\right]
\label{F1d}
\end{eqnarray}
with $a=b=c=d=1$ and $c_1=1$.

Fig.1e plots a special molecule constructed by two solitons and a KAK bound state with
$U$ given by \eqref{F1d},
$a=b=c=d=1,$ and $c_1=\sqrt{2}$.

Fig.1f shows a particular SM constructed by one M-shape soliton (MSS) and two single peak solitons with \eqref{F1d},
 $a=b=c=d=1,$ and $c_1=2$ at time $t=0$.

\input epsf
\begin{figure}[htbp]
\centering
{\includegraphics[height=4.5cm,width=5.5cm]{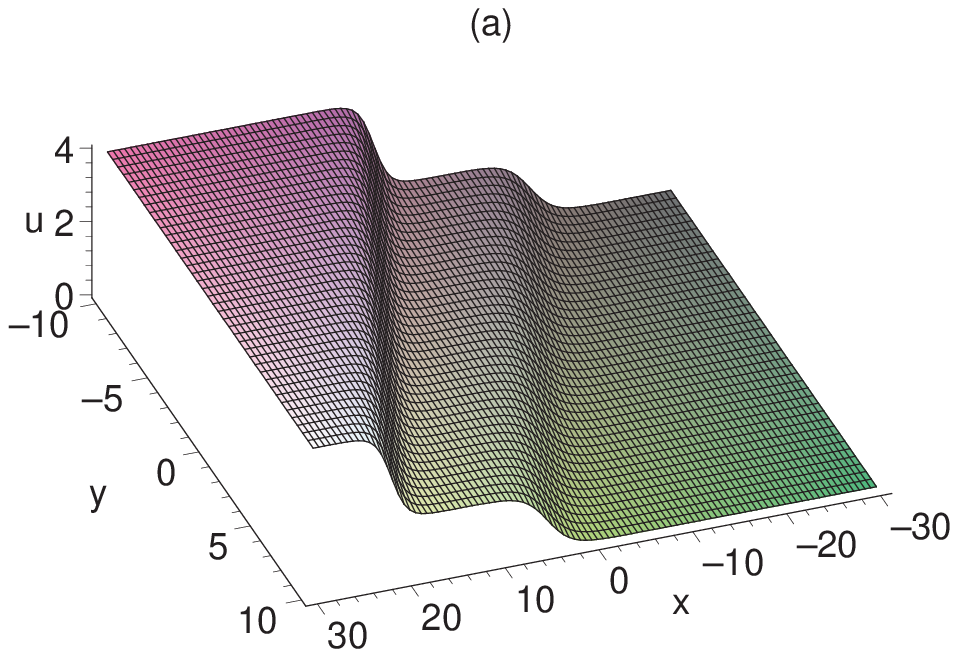}}
{\includegraphics[height=4.5cm,width=5.5cm]{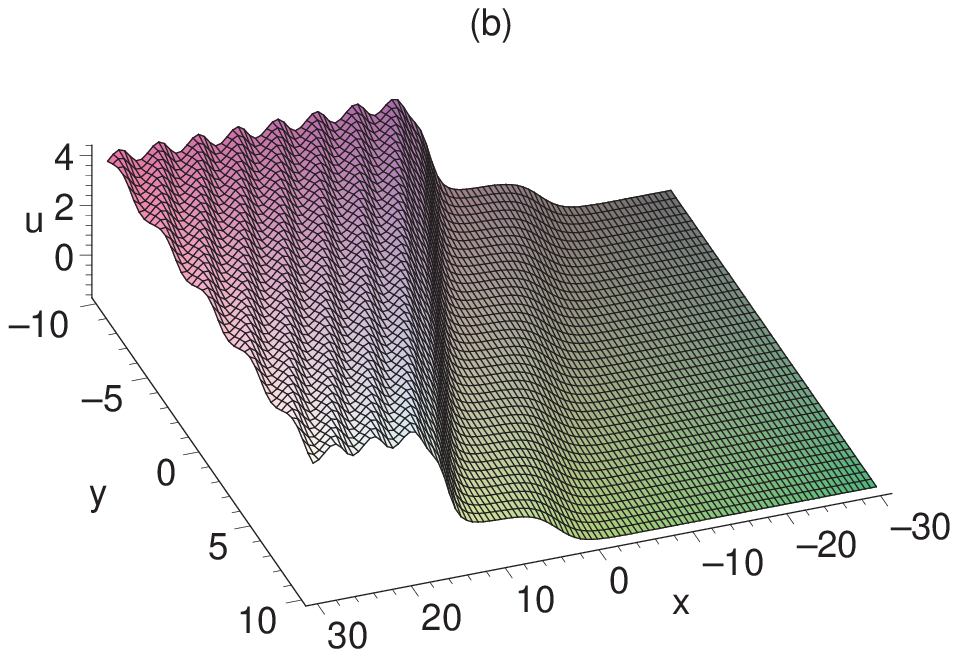}}
{\includegraphics[height=4.5cm,width=5.5cm]{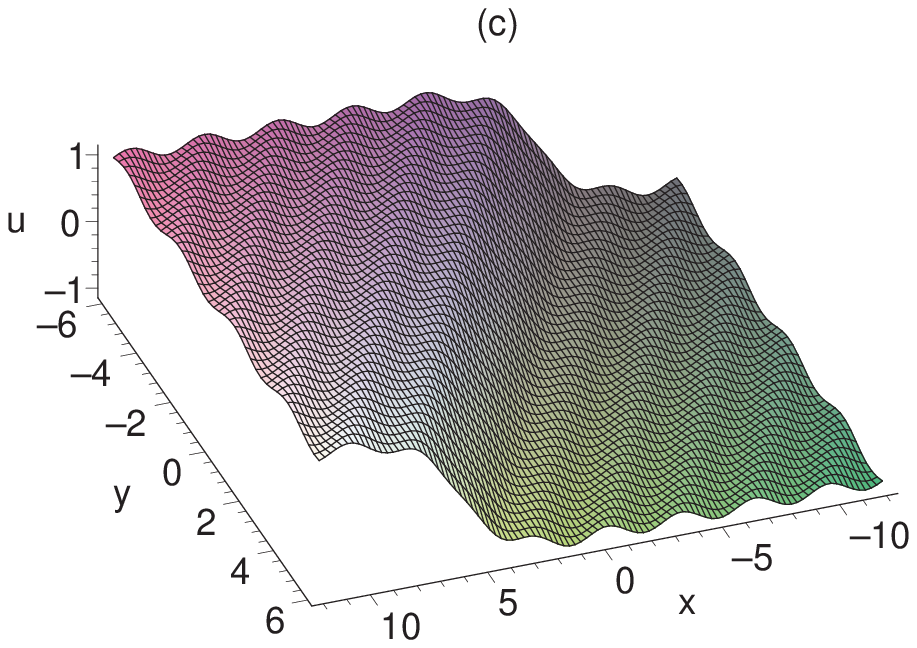}}
{\includegraphics[height=4.5cm,width=5.5cm]{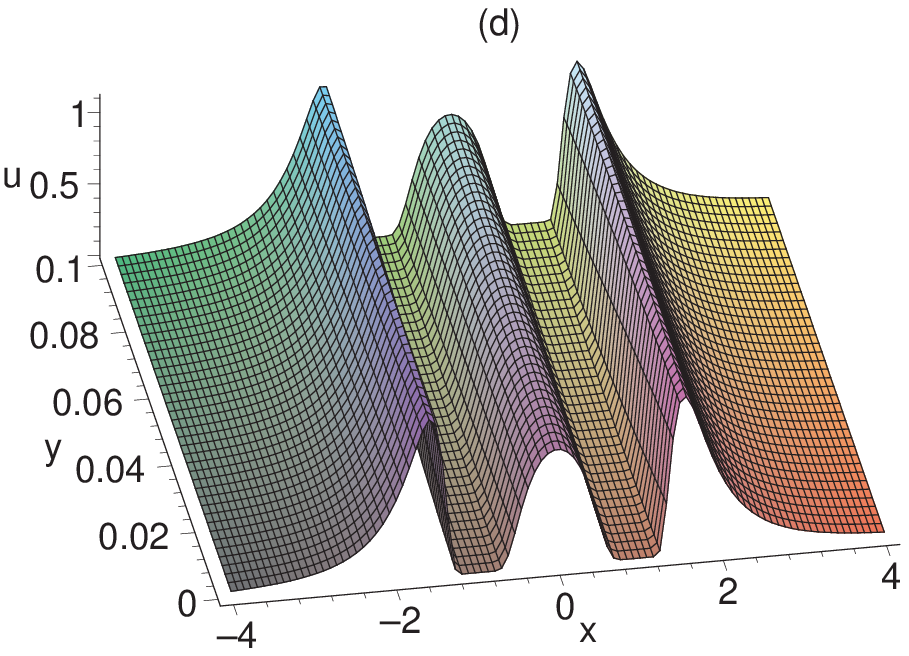}}
{\includegraphics[height=4.5cm,width=5.5cm]{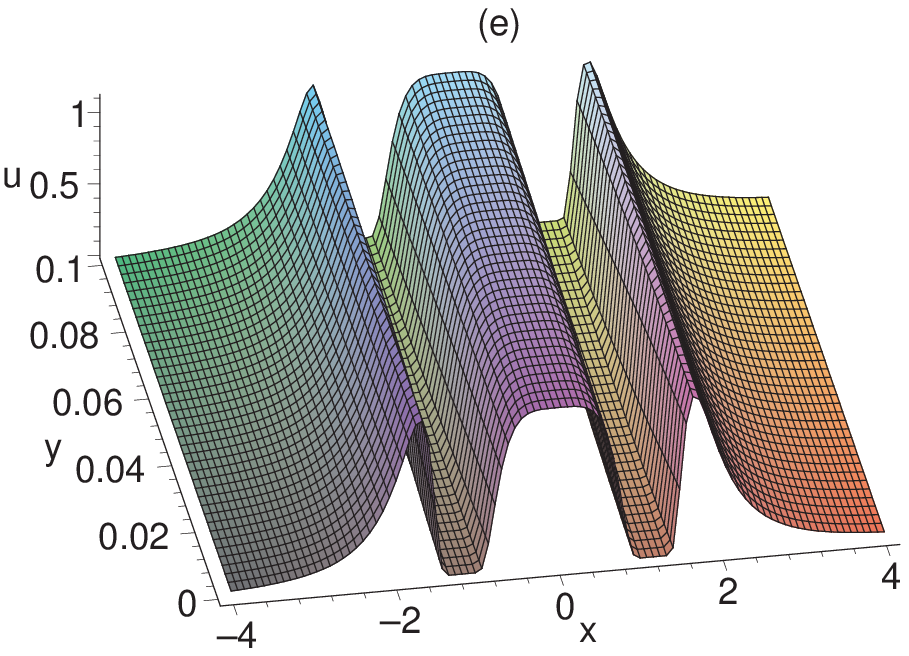}}
{\includegraphics[height=4.5cm,width=5.5cm]{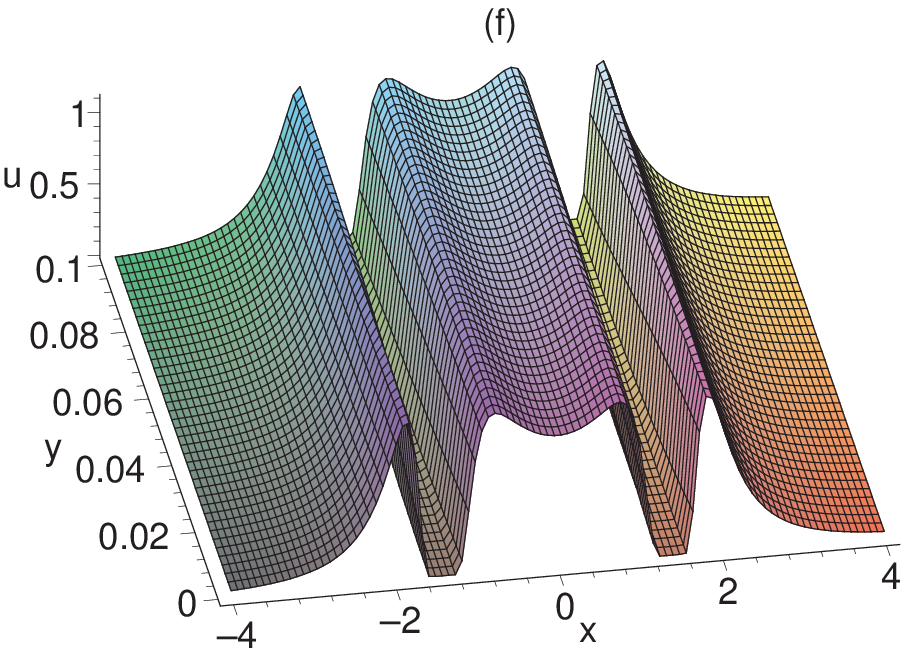}}
\caption{(a) Kink molecule (layered dislocation in solid physics) described by \eqref{F1a}. (b) HPK-kink molecule given by \eqref{F1b}. (c) Periodic kink expressed by \eqref{F1c}. (d) Three-soliton molecule presented by \eqref{F1d} with $c_1=1$. (e) Soliton-KAK-soliton molecule defined by \eqref{F1d} with $c_1=\sqrt{2}$. (f) Soliton-MS-soliton molecule determined by \eqref{F1d} with $c_1=2$.}\label{Fig1}
\end{figure}

Now a further interesting question is that can we find some special types of interaction solutions among some special D'Alembert type waves and other types of NNV waves?
To partially answer this question, we study the multiple soliton solutions of the NNV equation \eqref{NNV}.

To find multiple soliton solutions, Hirota's bilinear method can be used. By taking the transformations
\begin{equation}
u=2[\ln (f)]_{xy},\ v=2[\ln (f)]_{xx},\ w=2[\ln (f)]_{yy}, \label{uvwf}
\end{equation}
the NNV equation becomes
\begin{equation}
(f\partial_x-2f_x)D_y(2aD_x^3+2cD_x+D_t)f\cdot f
+(f\partial_y-2f_y)D_x(2bD_y^3+2dD_y+D_t)f\cdot f=0, \label{NNVf}
\end{equation}
where the Hirota's bilinear operator $D_{x_i},\ x_i=x,\ y,\ z,\ t$ is defined by
\begin{eqnarray}
D_{x_i}^n f\cdot g= \left.(\partial_{x_i}-\partial_{x_i'})^n f(x_i)g(x_i')\right|_{x_i'=x_i}. \label{Dx}
\end{eqnarray}
Eq. \eqref{NNVf} can be further bilinearized to
\begin{eqnarray}
&&D_y(2aD_x^3+2cD_x+D_t+D_z)f\cdot f=0,\label{NNVfy}\\
&&D_x(2bD_y^3+2dD_y+D_t-D_z)f\cdot f=0, \label{NNVfx}
\end{eqnarray}
by introducing an auxiliary variable $z$.

It is straightforward to write down the solutions of \eqref{NNVfy}--\eqref{NNVfx} ($\xi_i=k_ix+p_iy+q_iz+\omega_it+\xi_{i0}$)
\begin{eqnarray}
&&f=\sum_{\mu}\exp\left(\sum_{j=1}^N\mu_j\xi_j+\sum_{1\leq i<j}^N\mu_i\mu_j\theta_{ij}\right),\
\exp(\theta_{ij})=\frac{(p_i-p_j)(k_i-k_j)}{(p_i+p_j)(k_i+k_j)},
\label{exp}\\
&&\omega_i=-a k_i^3-b p_i^3-c k_i-d p_i,\ q_i= a k_i^3-b p_i^3+c k_i-d p_i,
\end{eqnarray}
or equivalently \cite{ABSR,ABJMP,full}
\begin{eqnarray}
f=\sum_{\nu}K_{\nu}\cosh\left(\frac12\sum_{i=1}^N\nu_i\xi_i\right), \ K_{\nu}=\prod_{1\leq i<j}^N\sqrt{(p_i-\nu_i\nu_jp_j)(k_i-\nu_i\nu_jk_j)},\label{cosh}
\end{eqnarray}
where the summation on $\mu=\{\mu_1,\ \ldots,\ \mu_N\}$ should be done for all permutations with $\mu_i=0,\ 1,\ i=1,\ \ldots,\ N$ and the summation on $\nu=\{\nu_1,\ \ldots,\ \nu_N\}$ should be done for all non-dual permutations with $\nu_i=1,\ -1,\ i=1,\ \ldots,\ N$. The permutations $\nu$ and $-\nu$ are called dual. In the solutions \eqref{exp} and/or \eqref{cosh}, the auxiliary parameter $z$ can be absorbed by the arbitrary constants $\xi_{i0}$.

Applying the velocity resonant mechanism \cite{LouS1,LouS2} to the multiple soliton solution \eqref{exp} of the NNV equation \eqref{NNV}, we have
\begin{eqnarray*}
\frac{k_{i}}{k_j}=\frac{p_i}{p_j}=\frac{a k_i^3+b p_i^3+c k_i+d p_i}{a k_j^3+b p_j^3+c k_j+d p_j},\ k_i\neq \pm k_j,\ p_i\neq \pm p_j,
\end{eqnarray*}
say,
\begin{eqnarray*}
p_l=-\sqrt[3]{\frac{a}b}k_l,\ l=i,\ j.
\end{eqnarray*}
If one considers the velocity resonant conditions for $n$ solitons, one can find the only solution is
\begin{eqnarray}
p_l=-\sqrt[3]{\frac{a}b}k_l,\ \xi_l=k_l\tau+\xi_{l0},\ l=1,\ 2,\ \ldots,\ n,  \label{vr}
\end{eqnarray}
where $\tau$ is just the travelling variable \eqref{tau1} of the D'Alembert type waves. In other words, the $n$-soliton molecule \eqref{uvwf} with \eqref{exp}, \eqref{vr} and $N=n$ is just a special D'Alembert type wave.

Now we consider special interactions among D'Alembert type waves and usual solitons. If a soliton molecule is constituted by $n$ solitons, then Eq. \eqref{exp} can be rewritten as
\begin{eqnarray}
f&=&\sum_{\mu}\exp\left(\sum_{j=1}^n\mu_j\xi_j+\sum_{1\leq i<j}^N\mu_i\mu_j\theta_{ij}\right)\exp\left(\sum_{j=n+1}^N\mu_j\xi_j\right)\\
&=&\sum_{\mu'}F_{\mu'}(\tau)\exp\left(\sum_{j=n+1}^N\mu_j\xi_j\right),\label{expsm}\\
 \xi_j&=&k_jx+p_jy-(a k_j^3+b p_j^3+c k_j+d p_j)t+\xi_{j0},\ p_j\neq -\sqrt[3]{\frac{a}b}k_j,\ n<j\leq N,\nonumber
\end{eqnarray}
\begin{eqnarray}
 F_{\mu'}(\tau)=\sum_{\mu''}\exp\left(\sum_{j=1}^n\mu_j\xi_j+\sum_{1\leq i<j}^N\mu_i\mu_j\theta_{ij}\right), \ \xi_j=k_j\tau +\xi_{j0},\ 1\leq j\leq n,\
\label{expU}
\end{eqnarray}
where the summation on $\mu'=\{\mu_{n+1},\ \ldots,\ \mu_N\}$ should be done for all permutations of $\mu_j=0,\ 1,\ n<j\leq N$, and the summation on $\mu''=\{\mu_{1},\ \ldots,\ \mu_n\}$ should be done for all permutations of $\mu_j=0,\ 1,\ 1\leq j\leq n$ for the fixed $\mu'$.

From the expressions \eqref{expsm}--\eqref{expU}, we know that $F_{\mu'}(\tau)$ is just a special D'Alembert type wave. The soliton molecule interaction solution \eqref{expsm} is an interaction solution among a special D'Alembert wave ($n$-soliton molecule) and $N-n$ solitons of the NNV equation \eqref{NNV}.

For $n=2,\ N=3$,  \eqref{expsm} becomes,
\begin{eqnarray}
&&f=F_1(\tau)+F_2(\tau)\mbox{\rm e}^{\xi_3},\ \xi_3=k_3x+p_3y-(a k_3^3+b p_3^3+c k_3+d p_3)t+\xi_{30},\label{f23}\\
&&F_1(\tau)\equiv 1+\mbox{\rm e}^{\xi_1}+\mbox{\rm e}^{\xi_2}+A_{12}\mbox{\rm e}^{\xi_1+\xi_2},\ \xi_j=k_j\tau +\xi_{j0},\ j=1,\ 2 \label{F1}\\
&&F_2(\tau)\equiv 1+A_{13}\mbox{\rm e}^{\xi_1}+A_{23}\mbox{\rm e}^{\xi_2}+A_{12}A_{13}A_{23}\mbox{\rm e}^{\xi_1+\xi_2},\label{F2}
\end{eqnarray}
where
\begin{eqnarray*}
&&A_{ij}=\frac{(p_i-p_j)(k_i- k_j)}{(p_i+p_j)(k_i+k_j)},\ 1\leq i<j\leq 3,\ \frac{p_1}{k_1}=\frac{p_2}{k_2}=-\sqrt[3]{\frac{a}b},\ \frac{p_3}{k_3}\neq -\sqrt[3]{\frac{a}b}.
\end{eqnarray*}
Fig. 2a displays the structure for the field $u$ described by \eqref{uvwf} with \eqref{f23} and the parameter selections
\begin{eqnarray}
&& a=b=c=1,\ d=2,\ k_1=-p_1=10,\ k_2=-p_2=8,\ k_3=7,\ p_3=\xi_{10}=-\xi_{20}=5,\ \xi_{30}=0.
\label{2aPara}
\end{eqnarray}
For $n=3,\ N=4$,  we have,
\begin{eqnarray}
&&f=F_1(\tau)+F_2(\tau)\mbox{\rm e}^{\xi_4},\ \xi_4=k_4x+p_4y-(a k_4^3+b p_4^3+c k_4+d p_4)t+\xi_{40},\label{f31}\\
&&F_1(\tau)\equiv 1+\sum_{j=1}^3\mbox{\rm e}^{\xi_j}+\sum_{1\leq i\leq j}^3A_{ij}\mbox{\rm e}^{\xi_i+\xi_j}+A_{12}A_{13}A_{23}\mbox{\rm e}^{\xi_1+\xi_2+\xi_3}, \label{F31}\\
&&F_2(\tau)\equiv 1+\sum_{j=1}^3A_{j4}\mbox{\rm e}^{\xi_j}+\sum_{1\leq i\leq j}^3A_{ij}A_{i4}A_{j4}\mbox{\rm e}^{\xi_i+\xi_j}+\prod_{1\leq i\leq j}^4A_{ij}\mbox{\rm e}^{\xi_1+\xi_2+\xi_3},\label{F32}
\end{eqnarray}
where $\xi_j=k_j\tau +\xi_{j0},\ j=1,\ 2,\ 3$.

Fig. 2b is a plot of the solution for the field $u$ described by \eqref{uvwf} with \eqref{f31} under the parameter conditions
\begin{eqnarray}
&& a=b=c=1,\ d=2,\ k_1=-p_1=10,\ k_2=-p_2=9,\ k_3=-p_3=8, \ k_4=7,\ p_4=5,\nonumber\\
&&
\xi_{10}=-\xi_{20}=7,\ \xi_{30}=\xi_{40}=0.
\label{2bPara}
\end{eqnarray}
For $n=2,\ N=4$,  we have,
\begin{eqnarray}
&&f=F_0(\tau)+F_1(\tau)\mbox{\rm e}^{\xi_3}+F_2(\tau)\mbox{\rm e}^{\xi_4}+F_3(\tau)\mbox{\rm e}^{\xi_3+\xi_4},\ \label{f22}\\
&&F_0(\tau)\equiv 1+\mbox{\rm e}^{\xi_1}+\mbox{\rm e}^{\xi_2}+A_{12}\mbox{\rm e}^{\xi_1+\xi_2},\\
&& F_1(\tau)\equiv 1+A_{13}\mbox{\rm e}^{\xi_1}+A_{23}\mbox{\rm e}^{\xi_2}+A_{12}A_{13}A_{23}\mbox{\rm e}^{\xi_1+\xi_2}\\
&&F_2(\tau)\equiv 1+A_{14}\mbox{\rm e}^{\xi_1}+A_{24}\mbox{\rm e}^{\xi_2}+A_{12}A_{14}A_{24}\mbox{\rm e}^{\xi_1+\xi_2}\\
&&F_3(\tau)\equiv A_{34}\left(1+A_{13}A_{14}\mbox{\rm e}^{\xi_1}+A_{23}A_{24}\mbox{\rm e}^{\xi_2}+A_{12}A_{13}A_{14}A_{23}A_{24}\mbox{\rm e}^{\xi_1+\xi_2}\right),
\end{eqnarray}
where $\xi_j=k_j\tau +\xi_{j0},\ j=1,\ 2,\ \xi_j=k_jx+p_jy-(a k_j^3+b p_j^3+c k_j+d p_j)t+\xi_{j0},\ j=3,\ 4$ and $\tau$ is given by \eqref{tau1}.

Fig. 2c shows the interaction solution among two separated solitons and one two-soliton bounded state for the field $u$ described by \eqref{uvwf} with \eqref{f22} and the parameter conditions
\begin{eqnarray}
&& a=b=c=1,\ d=2,\ k_1=-p_1=10,\ k_2=-p_2=9,\ k_3=8, \ k_4=7,\ \ p_3=6,\ p_4=2,\nonumber\\
&&
\xi_{10}=-\xi_{20}=7,\ \xi_{30}=\xi_{40}=0.
\label{2cPara}
\end{eqnarray}

\input epsf
\begin{figure}[htbp]
\centering
{\includegraphics[height=4.5cm,width=5.5cm]{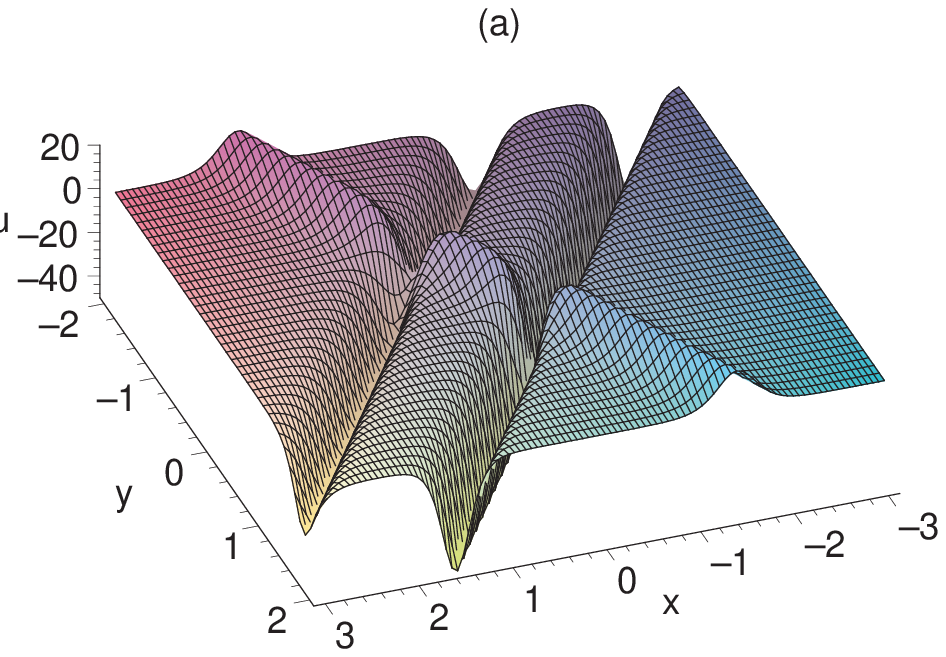}}
{\includegraphics[height=4.5cm,width=5.5cm]{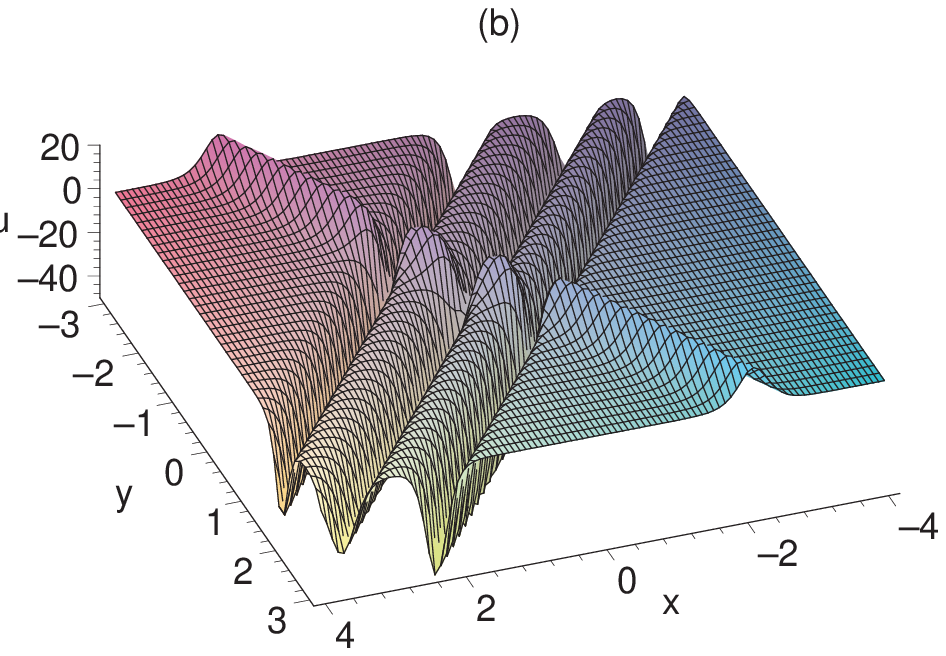}}
{\includegraphics[height=4.5cm,width=5.5cm]{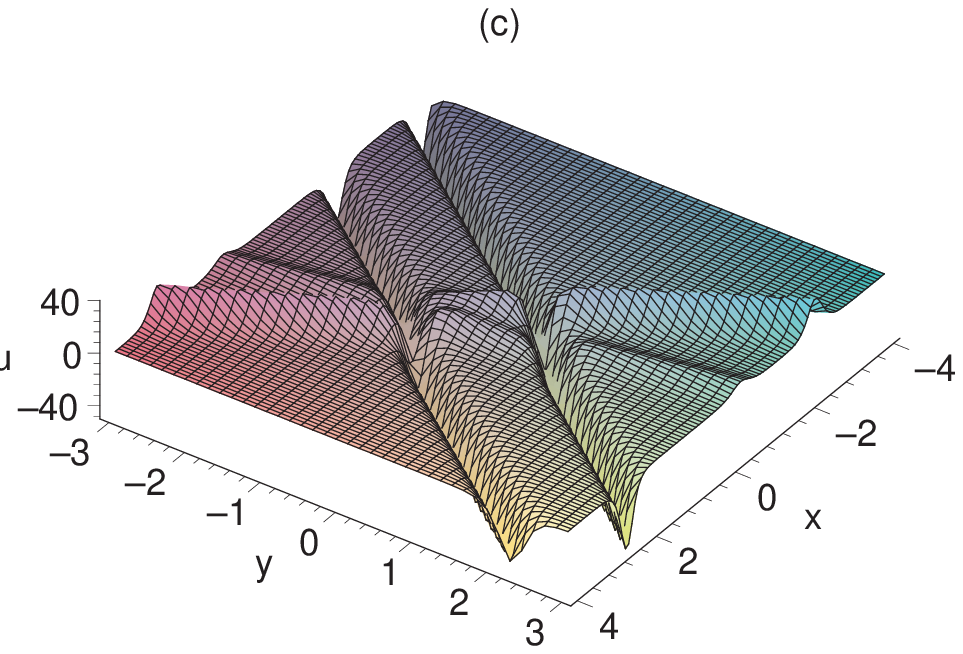}}
{\includegraphics[height=4.5cm,width=5.5cm]{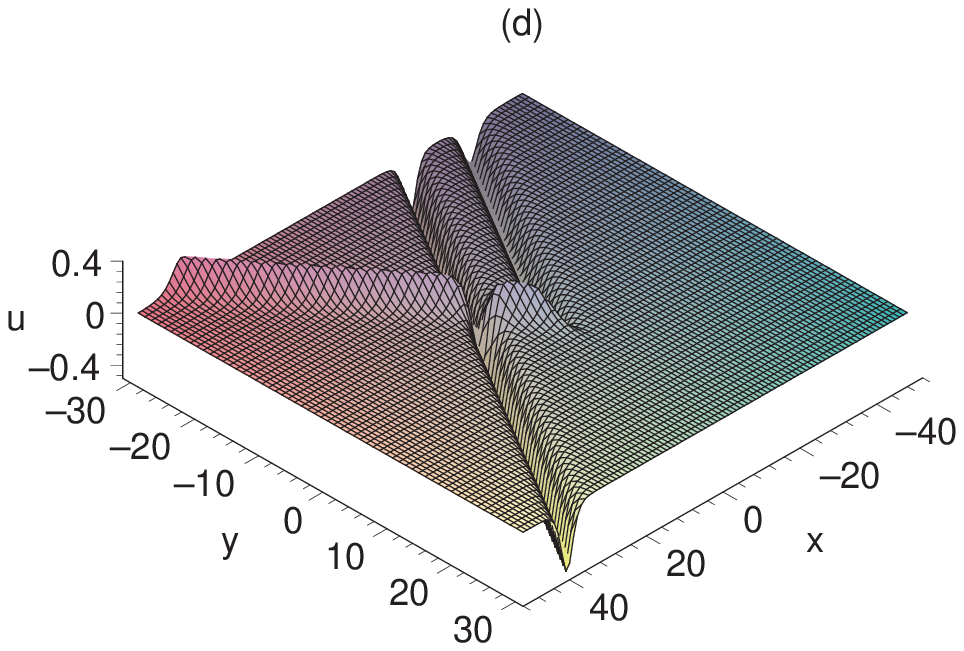}}
\caption{Interactions among solitons and special D'Alembert waves (soliton molecules) at time $t=0$: (a) interaction between a soliton and a two-soliton molecule, (b) interaction between a soliton and a three-soliton molecule, (c) interaction among two separated solitons and one two-soliton molecule, (d) interaction between a soliton and a soliton-solitoff molecule.}\label{Fig2}
\end{figure}
As a matter of fact, in addition to the interaction solution \eqref{expsm},
there are more abundant interaction structures among solitons and special D'Alembert type waves. Here, we just write down the general interaction form among one soliton and two D'Alembert type waves in the form
\begin{eqnarray}
&&f=F_1(\tau,\ z)+F_2(\tau,\ z)\mbox{\rm e}^{\xi},\ \xi=kx+py-(a k^3+b p^3+c k+d p)t+(ak^3-bp^3+ck-dp)z+\xi_{0},\label{f12xi}
\end{eqnarray}
where $ p\neq -\sqrt[3]{\frac{a}b}k,$ and $F_1(\tau,\ z)$ and $F_2(\tau,\ z)$ are the solutions of
\begin{eqnarray}
&&D_{\tau}\big[D_z-2abD_{\tau}^3-(d\sqrt[3]{a}+c\sqrt[3]{b})D_{\tau}\big]F_1\cdot F_1=0,\label{DF1} \\
&&D_{\tau}\big[D_z-2abD_{\tau}^3-(d\sqrt[3]{a}+c\sqrt[3]{b})D_{\tau}\big]F_2\cdot F_2=0,\label{DF2} \\
&&(\sqrt[3]{b}D_{\tau}D_z+b_1D_{\tau}+c_1\sqrt[3]{b}D_{\tau}^2-2b a_1\sqrt[3]{a^2}D_{\tau}^4-kD_z)F_1\cdot F_2=0,\label{F12a}\\
&&(\sqrt[3]{a}D_{\tau}D_z+b_2D_{\tau}+c_2\sqrt[3]{a}D_{\tau}^2-2a a_2\sqrt[3]{b^2}D_{\tau}^4+pD_z)F_1\cdot F_2=0,\label{F12b}
\end{eqnarray}
where $a_1=\sqrt[3]{a} \sqrt[3]{b}-k\sqrt[3]{a}+3p\sqrt[3]{b},$ $ a_2=\sqrt[3]{a} \sqrt[3]{b}-3k\sqrt[3]{a}+p\sqrt[3]{b},$ $c_1=6kp\sqrt[3]{a^2} \sqrt[3]{b^2}-6 bp^2\sqrt[3]{a}-d\sqrt[3]{a}-c\sqrt[3]{b}, $ $ c_2=6a\sqrt[3]{b}k^2-6kp\sqrt[3]{a^2}\sqrt[3]{b^2} +d\sqrt[3]{a}+c\sqrt[3]{b},$ $b_1=6 b k p^2\sqrt[3]{a}+d k\sqrt[3]{a}+ c k \sqrt[3]{b}$ and $b_2=-6 a k^2 p\sqrt[3]{b}-d p\sqrt[3]{a}- c p\sqrt[3]{b}$.

In addition to the solution \eqref{expsm} with $n=N-1$, the equation system \eqref{DF1}--\eqref{F12b} possesses many other types of special solutions. For instance,
\begin{eqnarray}
&&F_1=1+\mbox{\rm e}^{k_1\tau +q_1z+\tau_{10}}+\mbox{\rm e}^{k_2\tau +q_2z+\tau_{20}}+\frac{(k_1-k_2)^2}{(k_1+k_2)^2}\mbox{\rm e}^{(k_1+k_2)\tau +(q_1+q_2)z+\tau_{10}+\tau_{20}}, \label{2F1}\\
&&F_2=\mbox{\rm e}^{k_2\tau +q_2z+\tau_{20}}\left[1+\frac{(k_1-k_2)(\sqrt[3]{b}k_1-k)}{(k_1+k_2)(\sqrt[3]{b}k_1+k)}\mbox{\rm e}^{k_1\tau +q_1z+\tau_{10}}\right] \label{2F2}
\end{eqnarray}
with $q_{i}=2ab k_i^3+(d\sqrt[3]{a}+c\sqrt[3]{b})k_i,\ i=1,\ 2$ and $p=\sqrt[3]{a}k_2$.

Fig.2d shows a special interaction structure between a soliton and a special
D'Alembert type wave (a molecule constituted by a soliton and a solitoff) for the field $u$ described by \eqref{uvwf} with \eqref{f12xi}, \eqref{2F1} and \eqref{2F2} under the parameter selections
\begin{eqnarray}
&& a=b=c=k_1=1,\ d=2,\ k_2=0.8,\ k=0.6, \
\tau_{20}=-\tau_{10}=5,\ \xi_{0}=z=0
\label{2dPara}
\end{eqnarray}
at time $t=0$.

In summary, the missing D'Alembert type waves are discovered in a special nonlinear system, the NNV equation. The similar phenomena may be found for other types of nonlinear models \cite{KP34}. For the NNV equation, the $n$-soliton molecules are just the special type of D'Alembert waves. The interactions among solitons and the soliton molecules can be directly obtained from the multiple soliton solutions. It is found that there are other types of interactions among different types of D'Alembert waves and solitons. A special example, the interaction solution between a soliton-solitoff molecule and a separated soliton which can not be obtained from the multiple soliton solutions, is given explicitly. The D'Alembert type waves (including soliton molecules and soliton-solitoff molecules) are firstly found in nonlinear systems and deserve further investigated.

\section*{Acknowledgements\vspace{-0.2em}}
The work was sponsored by the National Natural Science Foundations of China (No.11975131) and K. C. Wong Magna Fund in Ningbo University.


\vspace{-1em}
\begin{thebibliography}{00}
\setlength{\itemsep}{-0.75ex}
\vspace{-0.5em}
\bibitem{KdV}D. G. Crighton,
Appl. Math. \bf 39, \rm 39 (1995).
\bibitem{Guo} H. Y. Guo, Z. H. Wang and K. Wu,
Phys. Lett. B \bf 264, \rm 277 (1991).
\bibitem{IST} C. S. Gardner, J. M. Greene, M. D. Kruskal and R. M. Miura,
Phys. Rev. Lett. \bf 19, \rm 1095 (1967).
\bibitem{Hirota}
R. Hirota, The direct method in soliton theory, Edited and translated by A. Nagai, J. Nimmo, C. Gilson, Cambridge Tracts
in Mathematics No. 155, Edition 1 (Cambrifge: Cambridge University Press) pp.1-61 (2004).
\bibitem{Gu} C. H. Gu, H. S. Hu and Z. X. Zhou, Darboux Transformations in Integrable Systems: Theory and their Applications to Geommetry,
Edition 1 (Dordrecht, Netherland: Springer) pp 1-64 (2005).
\bibitem{NNV1} L. P. Nizhnik, Sov. Phys. Dokl. \bf 25, \ \rm 706 (1980).
\bibitem{NNV2} A. P. Veselov, S. P. Novikov, Sov. Math. Dokl. \bf 30, \rm 588 (1984).
\bibitem{NNV3} S. P. Novikov, A. P. Veselov, Physica D \bf 18, \rm 267 (1986).
\bibitem{SP05}M. Stratmann, T. Pagel and F. Mitschke,
Phys. Rev. Lett. \bf 95,\rm 143902 (2005).
\bibitem{HK17}G. Herink, F. Kurtz, B. Jalali, D. R. Solli and C. Ropers,
Science  \bf 356,\ \rm 50 (2017).
\bibitem{LXM}X. M. Liu, X. K. Yao and Y. D. Cui,
Phys. Rev. Lett. \bf 121,\ \rm 023905 (2018).
\bibitem{Nano} C. Wang, L. Wang, et al,
Nanotechnol. \bf 30,\ \rm 025204 (2019).
\bibitem{LNS} K. Lakomy, R. Nath and L. Santos,
Phys. Rev. A \bf 86, \rm 013610 (2012).
\bibitem{SA} J. S. Peng, S. Boscolo, Z. H. Zhao, H. P. Zeng,
Sci. Adv. \bf 5, \rm 1110 (2019).
\bibitem{CK03} L. C. Crasovan, Y. V. Kartashov,  D. Mihalache, L. Torner, Y. S. Kivshar, V. M. Perez-Garcia,
Phys. Rev. E \bf 67,\ \rm 046610 (2003).
\bibitem{YB11} C. Yin, N. G. Berloff, V. M. Perez-Garcia, D. Novoa, A. V. Carpentier, H. Michinel,
Phys. Rev. A \bf 83,\ \rm 051605(R) (2011).
\bibitem{YanZY}Z. Y. Yan, Chaos, Solitons and Fractals \bf 122 \rm 25 (2019).
\bibitem{LouS1}S. Y. Lou, Soliton molecules and asymmetric solitons in fuid systems via velocity resonance, arXiv: 1909.03399. (2019); J. Phys. Commun. \bf 4,\ \rm 04110 (2020).
\bibitem{LouS2}D. H. Xu and S. Y. Lou, Acta Phys. Sin. \bf 69,\ \rm 014208 (2020) (in Chinese).
\bibitem{KdV5}A. S. Fokas and Q. M. Liu, Phys. Rev. Lett. \bf 77,\ \rm 2347 (1996).
\bibitem{LiB}Z. Zhang, X. Y. Yang and B. Li, Appl. Math. Lett. \bf 103, \rm 106168 (2020).
\bibitem{XYT}C. J. Cui, X. Y. Tang and Y. J. Cui, Appl. Math. Lett. \bf 103, \rm 106109 (2020).
\bibitem{YZW}Z. W. Yan and S. Y. Lou, Soliton molecules in Sharma-Tasso-Olver-Burgers equation arXiv: 1912. 13324.nlin.PS (2019), Appl. Math. Lett. \bf 104,\ \rm 106271 (2020).
\bibitem{KP34}S. Y. Lou, A novel (2+1)-dimensional integrable KdV equation with peculiar solution structures, arXiv:2001.08571 [nlin.SI] (2020).
\bibitem{ABSR}S. Y. Lou, and F. Huang,
Sci. Rep. \bf 7 \rm (2017) 869.
\bibitem{ABJMP}S. Y. Lou,
J. Math. Phys. \bf 59 \rm (2018) 083507.
\bibitem{full}S. Y. Lou, Acta Phys. Sin. \bf 69,\ \rm 010503 (2020) (in Chinese).

\end{thebibliography}
\end{document}